\documentstyle[epsfig,12pt]{article}
\newcommand{\cu}[1]{\chi \raisebox{-1ex}[0pt]{\scriptsize{$#1$}}}

\begin{document}

\title{ {\bf Validity of the one-body current \\ 
for the calculation of form factors \\ 
in the point form of relativistic quantum mechanics }}
\author{B. Desplanques$^a$\thanks{{\it E-mail address:}  
desplanq@isn.in2p3.fr}, 
 L. Theu{\ss}l$^a$\thanks{{\it E-mail address:}  
lukas@isn.in2p3.fr} \\
$^a$Institut des Sciences Nucl\'eaires (UMR CNRS/IN2P3--UJF),  
\\ F-38026 
Grenoble Cedex, France \\
}

\maketitle

\begin{abstract}
Form factors are calculated in the point form of relativistic quantum 
mechanics for the lowest energy states of a system made of two scalar 
particles interacting via the exchange of a massless boson. They are 
compared to the exact results obtained by using solutions of the 
Bethe-Salpeter equation which are well known in this case (Wick-Cutkosky model). 
Deficiencies of the point-form approach together with the single-particle 
current are emphasised. They point to the contribution of two-body currents 
which are required in any case to fulfil current conservation.
\end{abstract} 
\noindent 
PACS numbers: 11.10 St, 13.40.Fn, 13.60 Rj

\noindent
Keywords: form factors, relativity,  Wick-Cutkosky model 

\newpage
\section{Introduction}
Calculations of form factors often retain in a first approximation the 
single-particle current (impulse approximation). In most approaches, this 
contribution has to be completed by at least two-body currents, 
especially to fulfil current conservation. This also holds for 
relativistic approaches where a covariant calculation of the 
single-particle current does not necessarily imply that physics 
is properly accounted for. For each relativistic approach, it is 
therefore important to test the degree of validity by comparing 
its predictions to a case where an exact calculation can be performed. 
Neglecting vertex and mass corrections, as usually done on the basis 
that they are partly incorporated in the physical inputs, such an exact  
calculation is generally believed to be provided by the Bethe-Salpeter 
equation with an appropriate interaction kernel.

A particular case of interest is the Wick-Cutkosky model 
\cite{WICK,CUTK}, where the Bethe-Salpeter equation \cite{BETH} is 
solved in the ladder approximation while the interaction kernel 
results from the exchange of a scalar zero-mass boson  between two 
distinguishable scalar particles. Solutions can be obtained 
relatively easily due to an extra hidden symmetry. Using the 
expressions of the  Bethe-Salpeter amplitudes, calculations of 
form factors can be performed for the lowest bound states that 
we intent to consider here. Contrary to other approaches, there 
is no need to add two-body currents. The contribution of the 
single-particle current is sufficient in the present model to 
ensure current conservation, which can be checked with the expressions 
of the matrix elements of the current\footnote{This result 
together with some of the expressions used in the present 
paper for the Wick-Cutkosky model are part of a work in preparation.  A 
preliminary presentation was made in \cite{DESP0}.}. This model was 
used by Karmanov and Smirnov as a test of the description of 
form factors in the light-front approach for systems composed of 
scalar particles with small binding energy \cite{KARM}. They used the 
non-relativistic expression,  $-\frac{\alpha^2}{4n^2}$, for the binding 
energy, which differs from the exact one, but this does not seem to 
affect their conclusion. Interestingly enough, the comparison of 
the exact calculation with the non-relativistic one does not show 
much difference up to $Q^2 \simeq 100 \,m^2$ where $m$ is the 
constituent mass. Beyond, relativistic corrections with a log 
character slowly begin to show up. 

We propose to make a similar test for the point form of relativistic  
quantum mechanics, which is one of the forms proposed by Dirac, 
beside the instant and the light-front forms \cite{DIRA}. This form, 
which is much less known than the other two, has been developed 
in \cite{KLIN0} and recently 
used for a calculation of form factors of the deuteron \cite{KLIN1} 
and the nucleon \cite{WAGE}. In both cases, the form factors 
decrease faster with $Q^2$ than the non-relativistic ones. In 
the first case, the discrepancy with experiment tends to increase 
while, in the other one, it almost vanishes. We will not extend in 
this letter on the questions that are raised by this last observation. We will 
only present a few results which, by themselves, are quite significant. 
A more complete analysis will be presented elsewhere \cite{DESP1}. 

In the present study, we calculate the ground-state ($l=0$) form 
factor as well as a transition form factor to the first radial 
excited state. Different couplings are considered, corresponding 
to states weakly, moderately and strongly bound ($\alpha= 1.0, \; 3.0 \;{\rm 
and} \; 2\pi$ in terms of the QED coupling). The last value 
implies a total zero mass of the system, which is an extreme 
(unphysical) case, nevertheless interesting to look at, too. 
Momentum transfers, $Q^2$, up to 10 times the constituent mass 
squared, $m^2$, are considered. These different cases will provide 
a sample of results which are significant enough to test the validity 
of the single-current approximation in the point-form approach 
and give insight on the results presented in refs. \cite{KLIN1,WAGE}.  
\section{Expression of the single-particle  matrix element in different 
approaches}

\begin{figure}[htb]
\begin{center}
\mbox{ \psfig{ file=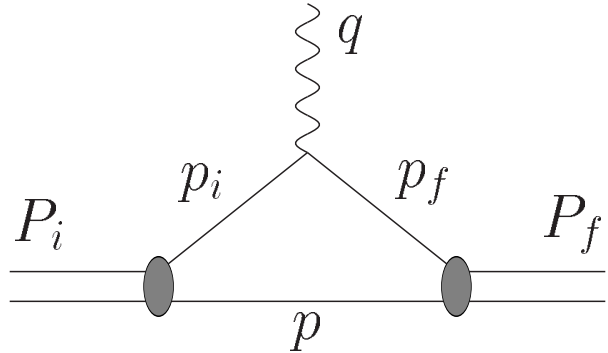, width=8cm}}
\end{center}
\caption{Representation of a virtual photon absorption on a two-body system 
with the kinematical definitions.}
\end{figure}  

The contribution which we are interested in is shown in Fig. 1. 
The general expression of the corresponding matrix element between 
two states with $l=0$, possibly different, is given by 
\begin{equation}
\sqrt{2E_f\,2E_i} \;\left<f|J^{\mu}|i\right> = F_1(q^2)\,(P^{\mu}_f+P^{\mu}_i) + 
F_2(q^2)\,q^{\mu},
\label{b1}
\end{equation}
where $q^{\mu}=P^{\mu}_f-P^{\mu}_i$. Current conservation imposes 
constraints on the form factors $F_1(q^2)$ and $F_2(q^2)$. For an 
elastic process,  $F_2(q^2)$ has to vanish but this result automatically 
stems from symmetry arguments alone. It does not imply that current 
conservation holds at the operator level, as it should. For an inelastic 
process, the following relationship has to be fulfilled: 
\begin{equation}
F_1(q^2)\,(M^2_f-M^2_i) + F_2(q^2)\,q^2= 0.
\label{b2}
\end{equation}

\noindent
{\bf Form factors using the Bethe-Salpeter amplitudes.\\}
For the model under consideration here, the general (and exact) 
expression of the matrix element of the current, which reduces 
to a single-particle one in this case, can be written in terms of the 
Bethe-Salpeter amplitudes, $\cu{P}( p )$,
\begin{eqnarray}
\sqrt{2E_f\,2E_i} \; \left<f|J^{\mu}|i\right> =    \hspace{9.5cm} \nonumber \\
i\int \frac{d^4p}{(2\pi)^4} \; \cu{P_f}  \Big(\frac{1}{2} P_f-p \Big) \;
\left( P^{\mu}_f+P^{\mu}_i-2\,p^{\mu}\right) 
\,(p^2-m^2)\; \cu{P_i}  \Big( \frac{1}{2} P_i-p \Big).
\label{b3}
\end{eqnarray}
For the Wick-Cutkosky model, the Bethe-Salpeter amplitudes take the 
form of a relatively simple integral representation, i.e. for the lowest energy 
state with a given orbital angular momentum $l$: 
\begin{equation}
 \cu{P}( p ) =\int_{-1}^1 dz \; \frac{g_n(z) \; Y_l^m(\hat{p}) \; 
|\vec{p}\,|^l}{(m^2 - 
\frac{1}{4}P^2- p^2-z\,P \cdot p-i\epsilon)^{n+2}},
\label{b4}
\end{equation}
with $n=l+1$. In this expression, $g_n(z)$ is the solution of a second order 
differential equation \cite{WICK,CUTK}, that can be solved easily. 

\noindent
{\bf Form factors in the non-relativistic limit.\\}
Opposite to the full relativistic calculation, a non-relativistic one can be 
performed. Using wave functions that are solutions of a Schr\"odinger equation, 
general expressions for both elastic and inelastic form factors can be obtained: 
\begin{eqnarray}
F_1(q^2)=  \int \frac{d \vec{p}}{(2\pi)^3} \;\; 
\phi_f\Big(\vec{p}-\frac{1}{4}\vec{q}\,\Big)
\;\; \phi_i\Big(\vec{p}+\frac{1}{4}\vec{q}\,\Big),\hspace{8mm} \nonumber \\
F_2(q^2) \, \frac{\vec{q}}{4m}=-\int \frac{d \vec{p}}{(2\pi)^3} 
\;\; \phi_f\Big(\vec{p}-\frac{1}{4}\vec{q}\,\Big) \;\; \frac{\vec{p}}{m} \;\;
\phi_i\Big(\vec{p}+\frac{1}{4}\vec{q}\,\Big).
\label{b11}
\end{eqnarray}
It can be checked that the above form factors verify the current conservation 
condition, Eq. (\ref{b2}), provided the interaction is local. The second 
form factor vanishes in the elastic case.

\noindent
{\bf Form factors in the point-form approach.\\}
It has been shown \cite{KLIN0} that a calculation of form factors 
in the point form approach could be performed relatively easily by 
using standard wave functions obtained from a mass operator of the form
\begin{equation}
M=M_{free}+M_{int}.
\label{b7}
\end{equation}
This includes a large class of wave functions, since  the sum 
of the kinetic and potential energies appearing in the standard 
Schr\"odinger equation can be identified with the operator $M^2$ (up 
to a factor). This holds for a two-body system and provided the 
energy $E$ is appropriately redefined. Therefore, wave functions entering 
the non-relativistic expressions of Eq. (\ref{b11}) can be used. How the matrix 
element of the single-particle current is calculated has been described 
in ref. \cite{KLIN1}. Instead of using an expression where appropriate 
boosts have to be performed, we here give one whose 
Lorentz-covariance is explicit: 
\begin{eqnarray}
\sqrt{2E_f\,2E_i} \; \left<f|J^{\mu}|i\right> = \sqrt{2M_f\,2M_i} 
\frac{1}{(2\pi)^3 }
 \hspace{6cm} \nonumber  \\
 \times \int d^4p \, d^4p_f \,  d^4p_i \, d\eta_f \, d\eta_i \; 
\delta(p^2-m^2) \;  \delta(p^2_f-m^2) \; \delta(p^2_i-m^2) \hspace{1.5cm} 
\nonumber \\ 
\times \theta( \lambda_f \cdot p_f) \; \theta(\lambda_f \cdot p) 
\;\theta(\lambda_i \cdot p)\; 
\theta(\lambda_i \cdot p_i) \; \delta^4(p_f+p-\lambda_f \eta_f) 
\;\delta^4(p_i+p-\lambda_i \eta_i)  \nonumber \\
\times \phi_f \Big((\frac{p_f-p}{2})^2\Big)  \;  
\phi_i \Big( (\frac{p_i-p}{2})^2 \Big) \;
\sqrt{(p_f+p)^2 \, (p_i+p)^2 } \;\;\;(p_f^{\mu}+ p_i^{\mu}) . 
\label{b8}
\end{eqnarray}
In this expression, all quantities are Lorentz-invariant ones (except obviously 
for the current which behaves as a 4-vector). The auxiliary variables, $\eta_i$ 
and  $\eta_f$, have been introduced to make the covariance manifest. When 
these variables are integrated over, they give rise to 3-dimensional 
$\delta(...)$-functions 
$$ 
\delta\Big(\vec{p}_i+\vec{p}-\frac{\vec{\lambda}_i}{\lambda_i^0}(p^0_i+p^0)\Big) 
\;\; {\rm and} \;\; 
\delta\Big(\vec{p}_f+\vec{p}-\frac{\vec{\lambda}_f}{\lambda_f^0}(p^0_f+p^0)\Big)
,$$
that are essential relations in the point-form approach. The $\lambda^{\mu}_i 
\; {\rm and} \; \lambda^{\mu}_f$ 4-vectors are unit vectors proportional to the 
4-momenta of the total system in the initial and final states, $\lambda^{\mu}_i 
= \frac{P^{\mu}_i}{M_i} \; {\rm and} \; \lambda^{\mu}_f= \frac{P^{\mu}_f}{M_f} 
$. They can be expressed in terms of the corresponding velocities\footnote{The 
quantity, $v$, here refers to the usual velocity 3-vector, $\vec{v}$, a notation 
that differs from the one employed in ref. \cite{KLIN0} where it represents a 
4-vector which corresponds to our 4-vector 
$\lambda^{\mu}$.}, 
$\lambda^0=(\sqrt{1-v^2}\,)^{\,-1}$ and $\vec{\lambda}= 
\vec{v}\,(\sqrt{1-v^2}\,)^{\,-1}$ (c=1).
 In the c.m., it can be checked that the 
wave function $\phi(...)$ only depends on the relative momentum of the two 
particles. On the other hand, it can also be verified, by direct integration or 
after performing a change of variable, that the current of the system under 
consideration is given by $(\, \left<J^0\right>\,,\,\vec{ \left<J \, \right>\,}) 
 = (\,1\,, \, \vec{v}\,)$, in agreement 
with the standard normalisation of the wave function 
$$ \int\frac{d\vec{p}}{(2\pi)^3} \;\phi^2(\vec{p}\,)=1.$$ 
Expressions of the form factors, $F_1(q^2)$ and $F_2(q^2)$, can be 
obtained in any frame. With an appropriate change of variables, they can 
always be expressed in the forms they take in the Breit frame (here defined by  
$\vec{v}=\vec{v}_f=-\vec{v}_i$), where they may be simpler. Using auxiliary 
quantities, $\tilde{F}_1(q^2)$ and  $\tilde{F}_2(q^2)$, they read: 
\begin{eqnarray}
F_1(q^2) \;  \sqrt{2M_f\,2M_i}  =  \tilde{F}_1(q^2) \, (M_f+M_i) 
      - \tilde{F}_2(q^2) \, (M_f-M_i), 
\nonumber \\
F_2(q^2) \;  \sqrt{2M_f\, 2M_i}  = -\tilde{F}_1(q^2) \, (M_f-M_i) 
      + \tilde{F}_2(q^2)\, (M_f+M_i),
\label{b9}
\end{eqnarray}
with  
\begin{eqnarray}
\tilde{F}_1(q^2)= \frac{1+v^2}{\sqrt{1-v^2}} \int \frac{d \vec{p}}{(2\pi)^3} 
\;\phi_f(\vec{p}_{tf}) \;\phi_i(\vec{p}_{ti}), \nonumber \\
\tilde{F}_2(q^2) \; \vec{v}= -\frac{1+v^2}{\sqrt{1-v^2}}    
\int \frac{d \vec{p}}{(2\pi)^3} 
\; \phi_f(\vec{p}_{tf}) \;\;  \frac{\vec{p}}{e_p} \;\; 
\phi_i(\vec{p}_{ti}).\label{b10}
\end{eqnarray}
In the above equations, the velocity $\vec{v}$, defined in the Breit frame, is 
related to the momentum transfer by the relation 
$v^2=\frac{Q^2+(M_f-M_i)^2}{Q^2+(M_f+M_i)^2}$. The (Lorentz-) transformed 
momenta are defined as:
$ (p^x,p^y,p^z)_{tf}= (p^x,p^y, \frac{p^z-v \, e_p}{\sqrt{1-v^2}})$ and
$ (p^x,p^y,p^z)_{ti}= (p^x,p^y, \frac{p^z+v \, e_p}{\sqrt{1-v^2}})$, 
together with $e_p=\sqrt{m^2+\vec{p}^{\,2}}$.

\noindent
{\bf Wave functions and analytical results.\\}
For the wave functions of the ground and first excited states, $\phi(\vec{p}\,)$ 
and $\phi^{*}(\vec{p}\,)$, we use solutions obtained with a Coulomb-like  
potential, 
\begin{eqnarray}
\phi(\vec{p}\,)=\sqrt{4 \, \pi} \frac{4 \, 
\kappa^{5/2}}{(\kappa^2+\vec{p}^{\,2})^2} 
,\hspace{1cm} \nonumber\\
\phi^{*}(\vec{p}\,)= \sqrt{4\pi} 
\frac{8 \, 
\kappa^{*5/2}}{(\kappa^{*2}+\vec{p}^{\,2})^3}(\kappa^{*2}-\vec{p}^{\,2}),
\label{b12}
\end{eqnarray}
where $\kappa^2=m^2-\frac{1}{4}M^2$, the 
total mass $M$ being that one obtained from the Bethe-Salpeter equation for the 
ground state. It has  been shown \cite{AMGH} that the spectrum of the normal 
states for the  
Wick-Cutkosky model, which has the same degeneracy as the Coulomb potential, 
could be reproduced with a 3-dimensional equation  and an effective interaction 
of the Coulomb-type\footnote{The spectrum of normal states of the  
Wick-Cutkosky model is 
reproduced within a factor 2 by an equation of the type $ 
[4(m^2+\vec{p}^{\,2})-M^2] \;\phi(r)= 4m \; \frac{\alpha_{eff}}{r} \; \phi(r)$, 
even for $\alpha_{eff} \rightarrow \infty$.}.
Due to the appearance of $M^2$ in such an equation, the energy 
redefinition mentioned previously is not even needed in the present case. The 
wave functions of Eqs. (\ref{b12}) should therefore be a good zeroth order 
approximation for our study, including the extreme case $M^2=0$. Accordingly, we 
assume $\kappa^{*2} = 0.25\,\kappa^2=0.25\,(m\alpha_{eff}/2)^2$, which differs 
from the Bethe-Salpeter result, $\kappa^{*2}=m^2-\frac{1}{4}M^{*2}$, by a few 
percent. With these wave functions, the form factors can be calculated 
analytically. We checked that the main features evidenced by our results were 
insensitive to this choice by using a different (numerical) wave function 
\cite{DESP1}, more in the spirit of the point-form approach, i.e. obtained 
directly from the linear mass operator, Eq. (\ref{b7}).

In the non-relativistic case, the elastic and inelastic form factors are 
respectively given by:
\begin{eqnarray}
F_1(q^2)= \frac{\kappa^4}{( \kappa^2+Q^2/16)^2},\;\;\;\;\;\;F_2(q^2)= 0, 
\hspace{2cm} \nonumber \\
F_1(q^2)= -\sqrt{2}\, \frac{64 \,\kappa^4 \, Q^2}{(9 
\kappa^2+Q^2)^3},\;\;\;\;\;\;
F_2(q^2)=-\sqrt{2}\, \frac{192\, \kappa^6 }{(9 \kappa^2+Q^2)^3}.
\label{b15}
\end{eqnarray}
Taking into account that $M^2_f-M^2_i= 3 \, \kappa^2$, one can verify that the 
condition of current conservation, Eq. (\ref{b2}), is fulfilled.

In the point-form approach, the elastic form factor, written in a way that 
resembles the non-relativistic one, reads: 
\begin{equation}
F_1(q^2=-Q^2)=  \frac{ \kappa^4\,\left( 1+2 \frac{Q^2}{4M^2}\right) }{
\Big(\kappa^2+\frac{Q^2}{16\,\big(1+ \frac{Q^2}{4M^2}\big)}\Big)^2 \; 
\left(1+ \frac{Q^2}{4M^2}\right)^4 } 
, \;\;\;\;\;\; F_2(q^2=-Q^2)=0.
\label{b13}
\end{equation} 
Interestingly, the factor $1/(1+ \frac{Q^2}{4M^2}) $ which multiplies 
the quantity $\frac{Q^2}{16 }$ in the denominator of $F_1(q^2)$ is 
the same as the one sometimes introduced by hand in order to account for the 
Lorentz-contraction effect (see discussion in ref. \cite{FRIA}). 
Contradicting asymptotic results (in QCD for instance), the range 
of validity of this recipe is limited to small $Q^2$. Curiously, 
part of the extra factors also look like a Lorentz-contraction 
effect. 

As for the inelastic form factors for a transition from the ground- to 
the first radially excited state, they are most easily expressed in terms of the 
quantities, $\tilde{F}_1(q^2)$ and $\tilde{F}_2(q^2)$, which can also be 
calculated analytically: 
\begin{eqnarray}
\tilde{F}_1(q^2)= -\sqrt{2}\, \frac{64 \,\kappa^4 
\,v^2\,(16\,m^2-4\kappa^2(1-v^2))}{(9 
\kappa^2+ v^2\,(16\,m^2-10\kappa^2)  +v^4\kappa^2)^3}\,(1+v^2) \,(1-v^2)^3, 
\nonumber \\
\tilde{F}_2(q^2)=-\sqrt{2}\, \frac{64\,(3+v^2) \, \kappa^6 }{(9 \kappa^2+  
v^2\,(16\,m^2-10\kappa^2)  +v^4\kappa^2  )^3}\,(1+v^2) \,(1-v^2)^4,
\label{b14}
\end{eqnarray}
where $v^2$ is defined after Eq. (\ref{b10}).  These expressions generalise 
those for the non-relativistic case, Eq. (\ref{b15}) (it is reminded that in 
this limit $Q^2 =16\,v^2\,m^2$).  In contrast however, they do not allow us to 
fulfil current conservation, Eq. (\ref{b2}). 
\section{Results} 
  
\begin{table}[htb]
\caption{Elastic form factor, $F_1(q^2)$, for the ground-state. Units for $E$ 
and $\kappa$ are the constituent mass, $m$. Results 
are presented for different couplings and, for each of them, for different 
approaches: Bethe-Salpeter equation  ($B.S.$), non-relativistic calculation 
($N.R.$) and point-form approach ($P.F.$). The value of $\alpha$  referred to in 
the table corresponds to the Bethe-Salpeter equation while the coupling for the 
``non-relativistic'' model, $\alpha_{eff}$, is chosen in such a way to reproduce 
the binding energy for the ground state.}
\begin{center}
\begin{tabular}{lccccc}
\hline  
  $Q^2/m^2$             &   0.01   &   0.1  &  1.0   &  10.0 \\ [1.ex] \hline
                        &          &        &        &       \\
 $\alpha=1$             &         &        &        &         \\ [0.ex] 
 $ E=0.0842,\kappa^2=0.0824$ &    &        &        &         \\ [0.ex] 
 $B.S.$                & 0.984    & 0.856  & 0.309  &  0.137-01  \\ [0.ex] 
 $N.R.$                & 0.985    & 0.864  & 0.323  &  0.135-01  \\ [0.ex]
 $P.F.$                & 0.984    & 0.853  & 0.298  &  0.097-01  \\  [1.ex]    

 $\alpha=3$            &          &        &        &         \\ [0.ex] 
 $ E=0.432,\kappa^2=0.385$ &      &        &        &         \\ [0.ex] 
 $B.S.$                & 0.996    & 0.962  & 0.705  &  0.139  \\ [0.ex] 
 $N.R.$                & 0.996    & 0.968  & 0.740  &  0.145  \\ [0.ex]
 $P.F.$                & 0.994    & 0.948  & 0.620  &  0.056  \\  [1.ex]    

 $\alpha=2\pi$         &          &           &          &         \\ [0.ex] 
 $ E=2.0,\kappa^2=1.0$ &          &           &          &         \\ [0.ex] 
 $B.S.$                & 0.998    & 0.983     & 0.848    &  0.339  \\ [0.ex] 
 $N.R.$                & 0.997    & 0.987     & 0.884    &  0.378  \\ [0.ex]
 $P.F.\,(E=1.90)$        & 0.613    & 0.397-01  & 0.110-03 &  0.124-06 \\ [0.ex]    
 $P.F.\,(E=1.95)$        & 0.181    & 0.142-02  & 0.191-05 &  0.196-08 \\ [1.ex]    
 
\hline
\end{tabular}
\end{center}
\end{table}

In Table 1, results are presented  for elastic form factors corresponding 
to small ($\alpha=1$), moderate ($\alpha=3$) and strong binding 
($\alpha=2\pi$). In the last case, the elastic form factors in the 
point form vanish at finite $Q^2$ and results are actually those 
obtained when approaching the limit $\alpha \rightarrow 2\pi$, 
with $E=1.90\,m$ and $E=1.95\,m$. For these energy values, the non-relativistic 
results, to which the previous ones may be compared, are 
essentially the same as for $E=2.0\,m$, given in the table. 

One immediately notices that the 
non-relativistic  results are very close to the exact ones for 
all cases, including the extreme case where the total mass of the 
system is zero. This agreement extends up to $Q^2=100\,m^2$ 
with an error of 20\% for $\alpha=1$ and 50\% for $\alpha=2\,\pi$. The 
discrepancy at small momentum transfers is typically of the 
order of $\frac{Q^2}{16m^2}$. Most probably, it can be traced 
back to the wave function used in the non-relativistic calculation 
or to the electromagnetic single-particle current. These 
ingredients do not fully account for corrections due to 
factors $\frac{m}{e}$ (in the potential for instance) or for the 
field-theory character of the Wick-Cutkosky model. A major lesson 
of these results is that relativistic effects are not necessarily 
important and that most probably, for any non-relativistic 
calculation of form factors, there exists a covariant 
calculation (the exact one in the present case) which gives very 
close results over a large range of momentum transfers. However, 
this does not mean that this calculation involves all relativistic 
effects and is physically relevant, especially with respect to current 
conservation. An example is provided by the deuteron 
electro-disintegration near threshold in the light-front approach. 
Two covariant calculations of the transition form factors have 
been shown to be very close to the non-relativistic ones up 
to $Q^2=10$ (GeV/c)$^2$ \cite{KEIS}, completely missing the contribution 
due to the pair term whose relativistic character is well known. 
In the non-relativistic approach, this contribution, which is 
required to fulfil current conservation, is essential to 
account for experiment in the low momentum transfer range. How 
this contribution  appears in the covariant light-front formalism 
was shown later on \cite{DESP2}.

The comparison with the point-form results evidences a discrepancy 
that increases with the momentum transfer as well as with the 
coupling strength. It becomes especially large when approaching 
the extreme case where $M=0$. Two features are worthwhile to be 
mentioned, that stem from examining the analytic expressions of 
the form factors, Eq. (\ref{b13}). First, there is a contribution 
to the squared-charge radius which varies like $\frac{1}{M^2}$, 
as it was found numerically in ref. \cite{WAGE}. Second, the 
form factor drops more quickly with $Q^2$ than in the exact 
or in the non-relativistic calculations, roughly like $\frac{1}{Q^6}$ instead 
of $\frac{1}{Q^4}$. This can be seen in Table 1 for $E=1.90\,m$ and $E=1.95\,m$ 
and is in agreement with what the examination of 
the Born amplitude reveals.  The effect becomes especially sizeable when 
approaching  $Q^2=4\,M^2$ and beyond.

\begin{table}[htb]
\caption{Inelastic form factors, $F_1(q^2)\; {\rm and}\; F_2(q^2)$, 
for a $l=0 \rightarrow l=0^{*}$ transition, and   $\alpha=3, 
E_i=0.4322\,m,\kappa_i^2=0.385\,m^2,E_f=0.1036\,m,\kappa_f^2=0.101\,m^2$ for the 
exact calculation and $0.096\,m^2$ for the non-relativistic one.}
\begin{center}
\begin{tabular}{lccccc}
\hline  
  $Q^2/m^2$             &   0.01   &   0.1  &  1.0   &  10.0 \\ [1.ex] \hline
                        &          &        &        &       \\
 $B.S.$ &   &   &        &         \\ [0.ex] 
 $F_1$                & -0.032-01    & -0.298-01 & -0.145-00   & -0.584-01 \\ 
[0.ex] 
 $F_2$                &  -0.369-00   & -0.340-00 & -0.165-00   & -0.665-02 \\ 
[1.ex]

 $N.R.$   &   &   &        &         \\ [0.ex] 
 $F_1$                & -0.032-01    & -0.296-01  & -0.151-00  &  -0.550-01 \\ 
[0.ex] 
 $F_2$                & -0.369-00    & -0.342-00  & -0.174-00  &  -0.636-02 \\ 
[1.ex]

 $P.F.$   &   &   &        &         \\ [0.ex] 
 $F_1$                & -0.101-01    &  -0.372-01 & -0.140-00  & -0.283-01  \\ 
[0.ex] 
 $F_2$                & -0.324-00    &  -0.293-00 & -0.119-00  & -0.022-02  \\ 
[1.ex]

\hline
\end{tabular}
\end{center}
\end{table}

Results for an inelastic transition are given in Table 2. The results 
obtained in the non-relativistic approach compare well with the exact 
ones over the full range of $Q^2$. The slight discrepancies are 
quite similar to those observed for the elastic form factors. While 
the point-form results compare well with the other ones in the 
intermediate range, $0.1<Q^2/m^2<3$, they fail at low and at 
large $Q^2$. In the former case, $F_1(q^2)$ does not go to zero when 
$Q^2\rightarrow 0$, violating the current 
conservation, Eq. (\ref{b2}). In the latter case, $F_2(q^2)$ 
evidences a change in sign around $Q^2 \simeq 10\,m^2$, also preventing 
one from fulfilling this relation. This can be traced back to the 
different behaviour of the intermediate form factors $\tilde{F}_1$ 
and $\tilde{F}_2$, which keep the same sign but scale like $Q^{-6}$ 
and $Q^{-8}$ at high $Q^2$, respectively. These results for an 
inelastic transition complement those for the elastic case. 
Again, a large discrepancy with exact results appears, but the 
bad behaviour of form factors at low as well as large momentum transfers 
is more clearly correlated with a violation of current conservation. 
\section{Discussion and conclusion}
The present study was motivated by the necessity to check the 
reliability of retaining the single-particle current for the 
calculation of form factors in the point form of quantum relativistic 
mechanics, which was recently employed in different works 
\cite{KLIN1,WAGE}. With this aim, we considered a simple model 
where the exact result is known. It was found that the non-relativistic 
approach does particularly well up to momentum transfers of 3-4 times 
the constituent mass, including the case of a strong binding. While the 
point-form approach does correctly for small bindings and small 
couplings, large discrepancies with the exact results appear as 
soon as the momentum transfer or the coupling increases. Detailed 
examination evidences three features. \\
- The form factors decrease more rapidly than they should. 
This points to an extra $\frac{1}{Q^2}$ 
dependence that is absent in the exact calculation. \\
- For strong couplings, corresponding to a sizeable presence of 
high momentum components in the wave function, the charge radius 
turns out to be much larger than the exact one. \\
- Finally, as emphasised by the results for an inelastic 
transition, current conservation is strongly violated. 

Although it is not quite certain, there is good reason to believe 
that the non-relativistic calculation does relatively well because 
it fulfils current conservation. With this respect, the failure 
of the point-form approach is likely to reside in the incomplete 
character of the current operator. Current conservation may be 
enforced by the replacement 
$$ \left<J^{\mu}\right> \;\;\rightarrow\;\; \left<J^{\mu}\right>- 
q^{\mu}\left<J_{\nu}\right> \cdot \,q^{\nu}/q^2.$$ 
Apart from the unsatisfactory character of this recipe, due to the 
presence of a pole at $q^2=0$ , it does not solve the problem of 
the too fast drop-off of the elastic charge 
form factor at high $Q^2$. Most probably, two-body currents have to be 
considered. This is consistent with the known fact that, in the point 
form approach, the interaction appears in all the components of the  
4-momentum operator \cite{KLIN2}. Whether a minimal set of two-body 
currents will be sufficient to provide results in better agreement 
with the exact ones is not clear however. These currents should also 
correct for the failure of the point-form results to reproduce the Born 
amplitude expected from the underlying field-theory model. 

The above results cannot be applied directly to the calculation of 
the deuteron or nucleon form factors \cite{KLIN1,WAGE}. However, 
the qualitative similarity in the results  strongly suggests that the 
kinematical boost, which provided a nice description of the nucleon form 
factors, represents an incomplete account of relativistic effects. More likely, 
the agreement is the consequence of neglecting significant contributions to the 
current. This conclusion is to be preferred with two respects. \\
- It leaves some room for the well known contribution of the coupling 
of the nucleon to the photon through vector-meson exchange (vector-meson 
dominance mechanism), which roughly provides half of the proton's squared 
charge radius. \\ 
- On the other hand, it leaves room for another relativistic effect related to 
the nature of the coupling of the constituents to the exchanged boson. This 
effect increases the form factor at high  $Q^2$ rather than the opposite as in 
ref. \cite{WAGE}. It is known to explain the different asymptotic 
form factors in a non-relativistic and a relativistic calculation 
in QCD, which scale like $\frac{1}{Q^8}$ and $\frac{1}{Q^4}$, respectively 
(see for instance refs. \cite{ALAB,CANO}).

Altogether, two-body currents should produce quite sizeable 
contributions. Their role seems to be more essential in the point form  
than in other approaches. 

{\bf Acknowledgements}
We are very indebted to W. Klink for information about his work, which largely 
motivated the present one. Helpful discussions with him are greatly 
appreciated.
\clearpage

\end{document}